\documentstyle[11pt,aaspp4]{article}

\include{floats}
\include{abbs}
\include{refs_journal}

\def\src{1055\,+\,018}

\def\NRAOcredit{The National Radio Astronomy Observatory is a facility
of the National Science Foundation, operated under a cooperative agreement
by Associated Universities, Inc.}

\received{1999 March 3}
\accepted{1999 April 20}
\journalid{518}{1999 June 20}
\paperid{995154}
\cpright{AAS}{1999}

\begin{document}

\title{ Radio Jet-Ambient Medium Interactions on Parsec Scales\\
in the Blazar \src}

\author{Joanne M. Attridge,\altaffilmark{1} David H. Roberts,\altaffilmark{2} and
John F. C. Wardle\altaffilmark{3}}
\affil{Department of Physics, MS-057, Brandeis University, Waltham,
MA 02454}

\centerline{\em Received 1999 March 3; accepted 1999 April 20; to be published in ApJL 1999 June 20}

\altaffiltext{1}{Current address: MIT Haystack Observatory, Route 40, Westford,
MA 01886-1299\\E-mail: jattridge@haystack.mit.edu}

\altaffiltext{2}{Visiting Scientist, National Radio Astronomy Observatory, and
Department of Astronomy and Jet Propulsion Laboratory, California Institute of
Technology\\E-mail: dhr@vlbi.astro.brandeis.edu}

\altaffiltext{3}{E-mail: jfcw@quasar.astro.brandeis.edu}

\begin{abstract}

As part of our study of the magnetic fields of active galactic nuclei, we have recently 
observed a large
sample of blazars with the Very Long Baseline Array. Here we report the discovery of
a striking two-component jet in the source \src\/ that consists of an inner spine
with a transverse magnetic field and a 
fragmentary but distinct boundary layer with a longitudinal magnetic
field. The polarization distribution in the spine strongly supports shocked-jet models,
while that in the boundary layer suggests interaction with the surrounding medium.
This behavior suggests a new way to understand the differing polarization properties of
strong- and weak-lined blazars. 

\end{abstract}


\section{Introduction} 
\label{s:intro}

Early VLBI polarimetry at 5~GHz revealed surprising differences between  strong-lined
and weak-lined AGN (loosely, ``quasars'' and ``BL Lacertae  objects''; 
Cawthorne et al.\ 1993).
Sources with prominent emission lines, such as the classic  quasars 3C\,273 and
3C\,345, are found to have primarily longitudinal magnetic  fields in their jets
(``$B_{\parallel}$''), while in weak-lined sources, such as  the BL Lacertae objects
OJ\,287 and BL Lac itself, the magnetic field is transverse to the jet axis
(``$B_{\perp}$''). However, this distinction is not universal, and sources with
$B_{\parallel}$ further out in their jets can also display $B_{\perp}$ closer to the
core (Lepp\"{a}nen, Zensus, \& Diamond 1995). In the  shocked relativistic jet 
paradigm under which these
facts are generally understood, knots with $B_{\perp}$ are the result of transverse
shocks that compress and partially order an initially tangled magnetic field
(Hughes, Aller, \& Aller 1985). Regions with $B_{\parallel}$ are thought 
to arise from interactions
between the jet and the ambient medium, in which shear of the jet flow stretches the
magnetic field along the jet (Wardle et al.\ 1994), or where the jet fluid is compressed
against its boundaries (Laing 1980). 

However successful, this picture does not explain why sources with different optical
characteristics exhibit different jet polarizations. In an attempt to better
characterize the relationship between optical and radio properties of active galactic nuclei,
we have
observed a sample of 45 blazars with the Very Long Baseline Array\footnote{\NRAOcredit} 
(VLBA) at 5~GHz at
several epochs (Attridge, Wardle, and Roberts, in preparation). Here we present
images of \src\/ (also known as J1058\,+\,0133, 4C\,01.28, DA\,293, and OL\,093), 
a blazar with $z = 0.889$
(Wills \& Lynds 1978). At this redshift 1~mas corresponds to $7.27 h^{-1}$~pc, and a proper
motion of 1~mas~yr$^{-1}$ corresponds to an apparent transverse speed of $\beta_{app}
= 44.7 h^{-1}$ ($H_0 = 70 h$~km~s$^{-1}$~Mpc$^{-1}$, $q_0 = 0.05$).  The rest-frame
emission-line equivalent widths of this source (Baldwin, Wampler, \& Gaskel 1989; 
Falomo, Scarpa, \& Bersanelli 1994),
$W_{eq}$(C~III]~$\lambda$1909)~$\simeq$~8~\AA\/, 
$W_{eq}$(Mg~II~$\lambda$2800)~$\simeq$~5--12~\AA\/, lie near the boundary 
$W_{eq} = 5$~\AA\ used by some authors to divide blazars into weak- and  strong-lined
categories (Morris et al.\ 1991).

\section{Observations}

Images of \src\ were made from data taken on 1997 January 24 (epoch 1997.07)
using all ten antennas of the VLBA plus a single antenna from the VLA
(``VLBA$+$Y1''). The data were recorded at 5~GHz ($\lambda=$6~cm) with a total
bandwidth of 64~MHz, and processed on the VLBA correlator. Approximately 4~hr
were spent on source, leading to a root mean square noise of
$\sim$75~$\mu\/$Jy~beam$^{-1}$. Calibration and hybrid imaging were done using
the NRAO AIPS and Caltech DIFMAP packages. In our notation, the total
intensity is $I$ and the linear polarization is represented by the complex quantity
$P = Q+iU = pe^{2i\chi} = mI e^{2i\chi}$, where $p = (Q^2+U^2)^{1/2} = mI$ is the
polarized intensity, $m$ is the fractional linear polarization, and $\chi$ is the
position angle of the electric vector. The integrated rotation measure of
\src\ is $-45$~rad~m$^{-2}$ (Kim, Tribble, \& Kronberg 1991), which we take to 
be Galactic. This
amounts to $\Delta \chi =  9^{\circ}$ at 5~GHz and has been removed in determining
the direction of the  magnetic field. Any additional Faraday rotation in or near the jet must
be unimportant at 5~GHz; otherwise, we would not find the striking magnetic field
configuration described below. Complete details of VLBI polarization calibration and
imaging techniques may be found in Cotton(1993), Roberts, Wardle, \& Brown (1994), and
Aaron (1996).

\section{Results}

In Figure ~1$a$ we show the total intensity distribution of \src\/.  This
naturally weighted image has a dynamic range (peak brightness/off-source 
noise) of $\sim$17,000:1. A typical core-jet source, \src\ contains a bright 
unresolved component to the east and a jet extending at least 35~mas to  the
west  northwest. The jet is resolved transverse to its axis (deconvolved
width ${_>\atop^{\sim}}$3~mas) and broadens considerably beyond $\sim$12~mas
from  the core. The  true nature of the jet is revealed by the linear
polarization distribution shown in Figure ~1$b$ (dynamic range $\sim$1100:1).
There we plot contours of linearly polarized intensity $p$ with ticks showing
the orientation $\theta=\chi+\frac{\pi}{2}$ of the jet magnetic field (assuming
optically-thin sychrotron radiation). It is
apparent that the jet consists of two distinct parts: (1) a spine lying along
the jet axis and containing a series of knots in which the magnetic field is
predominantly perpendicular to the axis, and (2) a fragmentary boundary layer
in which the magnetic field is predominantly parallel to the axis. Figure~2
superposes the $I$ and $P$ distributions and shows clearly that the
$B_{\parallel}$ regions lie on the outermost visible edges of the jet. A
natural interpretation of this two-component structure is that (1) transverse
shocks dominate within the jet spine, so the net field there is perpendicular
to the jet axis, and (2) in the boundary layer the magnetic field is
determined by the jet's interaction with the surrounding emission-line gas,
resulting in a longitudinal field. This configuration is reminiscent of the
``spine-shear  layer'' morphology seen in the kiloparsec-scale jets of sources such as 
3C\,31 (Laing 1996) and 3C\,353 (Swain, Bridle, \& Baum 1996). Apparently \src\ displays
simultaneously the magnetic field configurations characteristic of both weak- and
strong-lined blazars, and as such provides us with an opportunity to study both the
shocked-jet paradigm  and the relationship between magnetic field orientation and
optical properties for blazars.

In Figure~3 we show the fractional polarization of \src\/. At the jet  boundaries $m$
is quite high, exceeding $40\%$ in places, so the magnetic  field is well ordered,
both across the sky and along the line of sight. The peak polarization in the jet
spine is only $\sim$10\%, presumably due to a combination of partially tangled
magnetic field in the knots and beam dilution by the unshocked jet (Wardle et al.\ 1994).
There may also be some cancellation by any part of the boundary layer that intersects
the line of sight. The magnetic field in the jet spine is predominantly $B_{\perp}$,
so the shocked-jet model predicts the knots to be more highly polarized than the
regions between them (Wardle et al.\ 1994). That  this is so can be seen in  Figure~3, and
provides strong support for this model.

\section{Discussion}

There are some difficulties with the interpretation of the outer jet  in \src\
as  a parsec-scale version of the sheaths seen on much larger scales, i.e.,
as the  cylindrically symmetric outer layer of a coaxial jet. First, why is
the jet in \src\ so wide so close the core? Three possibilities suggest
themselves: (1) the true core is east of the easternmost component in our
images and invisible either because it points away from us or because it is severely
self-absorbed, (2) the jet initially expands very rapidly but is quickly
collimated, perhaps by the same forces that create the boundary layer, or (3)
the jet is initially pointing almost exactly at us, but bends away from the
line of sight within the first $\sim0.5$~mas. Second, why are the jet
boundaries visible on only one side of  the jet at a time? Perhaps they do
not form a complete ``sheath,'' but are those parts of the surface of the jet
where the interaction with the ambient  medium is strongest, most likely at
the outer edges of smooth bends or at  places where the jet is deflected. The
jet curves to the south within its first 10~mas and then turns northward
over the next  15~mas, which is consistent with this possibility.

On kiloparsec scales, \src\ is a Fanaroff-Riley Type~II radio source, with hot spots
almost due north and south of the core and a prominent jet in  position angle
$\sim180^{\circ}$ connecting the southern hot spot to the core; there is no sign of
a jet to the north (Murphy, Browne, \& Perley 1993). This source is thus an 
example of significant
misalignment ($\sim120^{\circ}$) of jet orientation between parsec and kiloparsec scales. From tapered
versions of Figure~1$a$ and from 1.6~GHz VLBI (Romney et al.\ 1984; Bondi et al.\ 1996) there are
suggestions that beyond $\sim$$35$~mas from the core the jet continues to bend to the
north. It would appear that on larger scales, the jet must bend through a very  large
angle (in projection) in order to connect with the kiloparsec-scale  jet to the
south. An alternative possibility is that the parsec-scale jet  continues to the
north and that the kiloparsec-scale jet to the south is  fed by an as-yet unseen
parsec-scale counterjet. This seems less likely, since it would require {\em
intrinsic} bends in both jets of more than $90^{\circ}$, assuming that the visible
parsec-scale jet is pointing nearly toward us. High dynamic range VLBA images at
lower frequencies, with superior surface brightness sensitivity and wider fields of
view, may  enable us to locate the bends and determine if and how the parsec- and 
kiloparsec-scale jets are connected.

Using the image presented here and two VLBA snapshot images at epochs 1996.02 and
1996.96, we have determined preliminary proper motions for the three brightest jet
spine components relative to the the easternmost component (the model fits are shown
in Table~1). Unfortunately, the time base is short. For components C1 and C2, we find
upper limits in apparent speed of about $5h^{-1}c$. For component C3, which is
closest to the core, we derive a provisional speed of $v_{app} = (12 \pm 4.5)
h^{-1}c$. Since the jet spine in \src\ has bright, well-defined knots, it will be
straightforward to obtain reliable apparent velocities and accelerations for all of
the knots over the next few years. This will also enable us to determine the physical
parameters of the shocks in the jet spine, and their evolution (Wardle et al.\ 1994).
Equally interesting  will be attempting to detect proper motions for the boundary
layer features. If this proves possible, the result would be a  direct measurement of
shear in a parsec-scale relativistic jet.

Velocity gradients across kiloparsec-scale jets have been inferred from their
polarimetric properties (Laing 1996; Swain et al.\ 1996). If similar gradients are common on
parsec scales,  they may provide a new way to understand the connection between the
apparent magnetic field orientation in blazar jets and the equivalent width of their
broad emission lines. We suggest that the spine and the sheath have significantly
different Lorentz factors $\gamma$: $\gamma_{spine}$ and $\gamma_{sheath}$, respectively, with
$\gamma_{spine} > \gamma_{sheath}$. The radiation from the faster-moving spine is
beamed into a cone of opening angle $\sim 1/\gamma_{spine}$, while that of the
sheath  is beamed into a wider cone of  opening angle $\sim 1/\gamma_{sheath}$. At
small angles to the line of sight, $\theta < \theta_{crit}$, the polarization image
is dominated by the faster moving spine, with its predominantly transverse magnetic
field. At $\theta > \theta_{crit}$, the polarization image is  dominated by the
sheath, with its longitudinal magnetic field. If we further assume that the optical
continuum radiation is also beamed with a Lorentz factor $\gamma_{opt} \ge
\gamma_{spine}$, then at small angles to the line of sight, the optical continuum is
strongly Doppler boosted with  respect to the line emission, and the  object will be
classified as a ``weak-lined blazar.'' At larger angles  to the line of sight, the
continuum will be far less prominent, and the  object will be classified as a
``broad-lined blazar'' (Antonucci 1993). Close to the core, where the sheath has not fully
developed, transverse magnetic fields are likely to be observed in both classes of
object (Lepp\"{a}nen et al.\ 1995).

We conclude that \src\ is a touchstone for a number of
important topics in the  physics of active galactic nuclei. Through it we can study the
formation of the jet-like structure both along and across the axis, including
bending and collimation, the conversion of bulk energy to microscopic
particle energy and radiation, and the formation, propagation, and decay of
shocks.

\acknowledgments
Radio astronomy at Brandeis University is supported by the National Science
Foundation. The National Radio Astronomy Observatory is  a facility of the
National Science Foundation, operated under a cooperative  agreement by
Associated Universities, Inc. We thank many colleagues for their questions and
suggestions on this work. D.~H.~R. thanks M.~Goss, M.~Cohen, S.~Kulkarni, and
R.~Preston for their hospitality.

\bigskip
Correspondence and requests for materials should be addressed to J.~M.~A.\\
(jattridge@haystack.mit.edu).

\clearpage

\begin{figure}
\plottwo{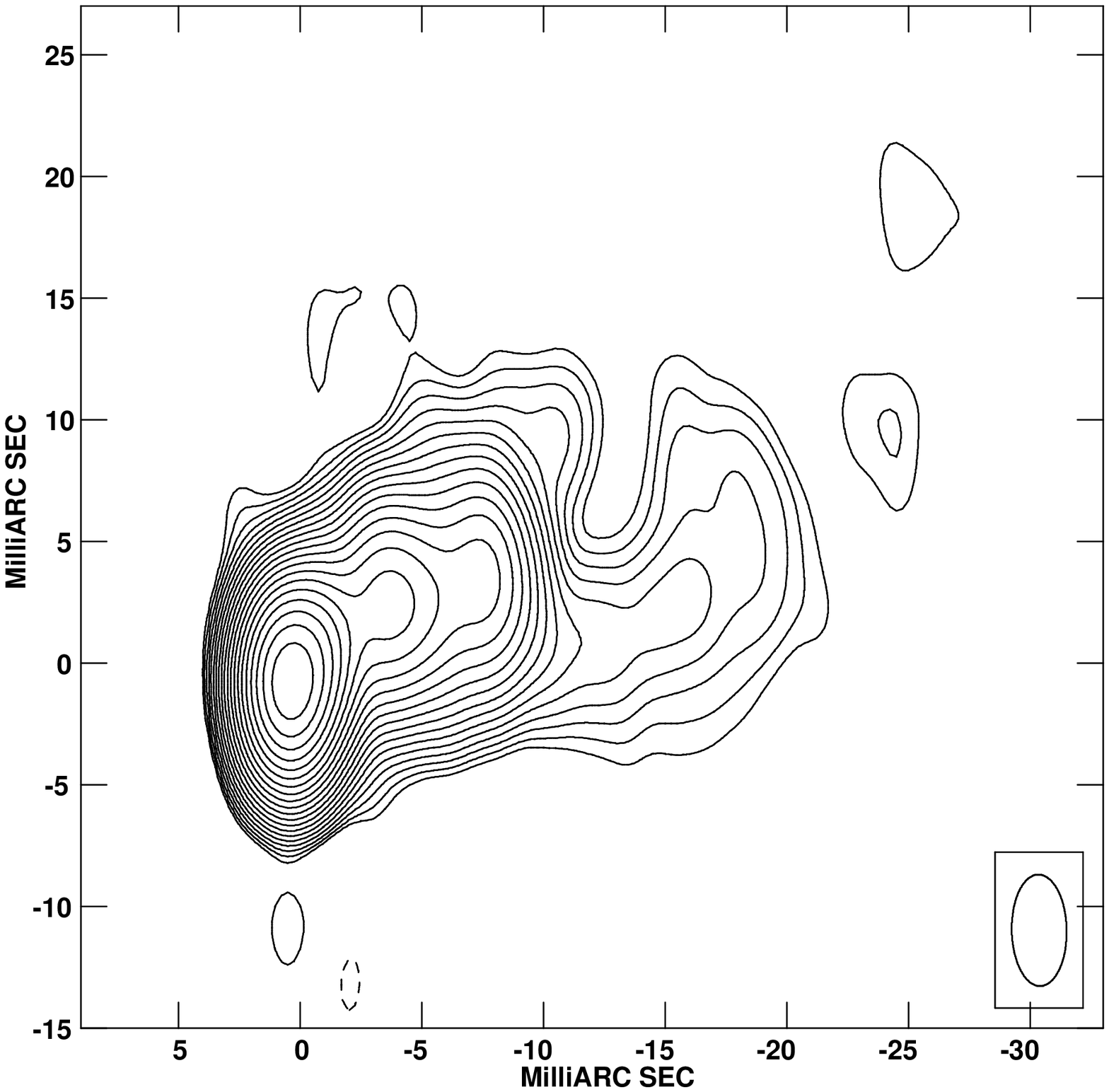}{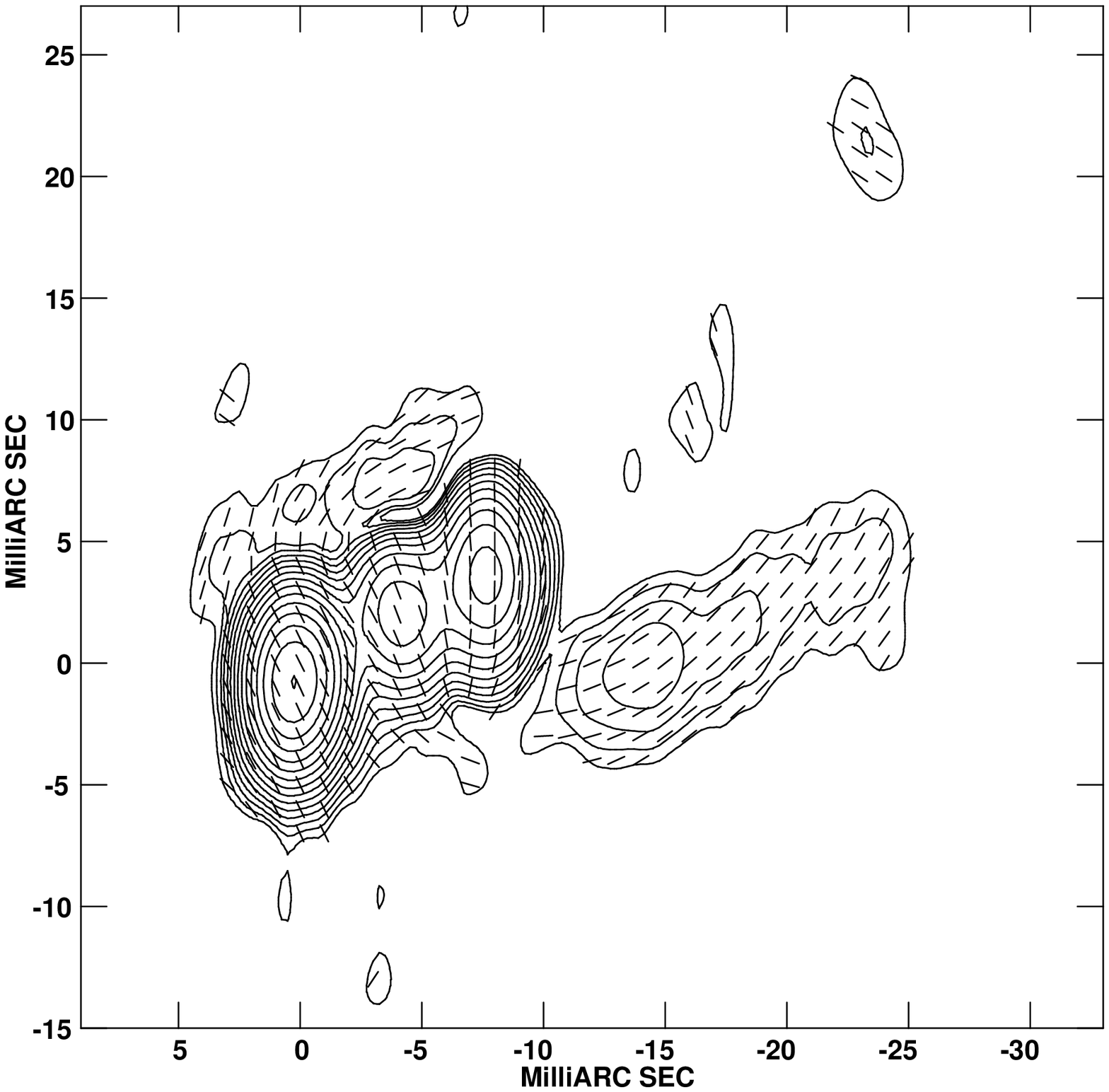}
\figcaption{Naturally weighted images of the blazar \src\/, epoch 1997.07,
made with the VLBA$+$Y1 at 5~GHz. ($a$) Total intensity distribution, with
contours of $I$ at $-1.75$, 1.75, 2.47, \ldots\ [factors of $\sqrt{2}$]
\ldots, 896, and 1270 mJy~beam$^{-1}$; the peak is 1720~mJy~beam$^{-1}$, and
the restoring beam shown in the lower right is 4.6~$\times$~2.3~mas 
at $\phi = 1^{\circ}$. ($b$) Linear
polarization distribution, with contours of $p$ at 0.5, 0.707, \ldots\
[factors of $\sqrt{2}$] \ldots, 45.3, and 64 mJy~beam$^{-1}$; the peak is
64.6~mJy~beam$^{-1}$. The ticks show the orientation $\theta = \chi +
\frac{\pi}{2}$ of the magnetic field in the source. The restoring beam is the
same as for $I$.\label{fig:1055I+P}}
\end{figure}

\clearpage

\begin{figure}
\plottwo{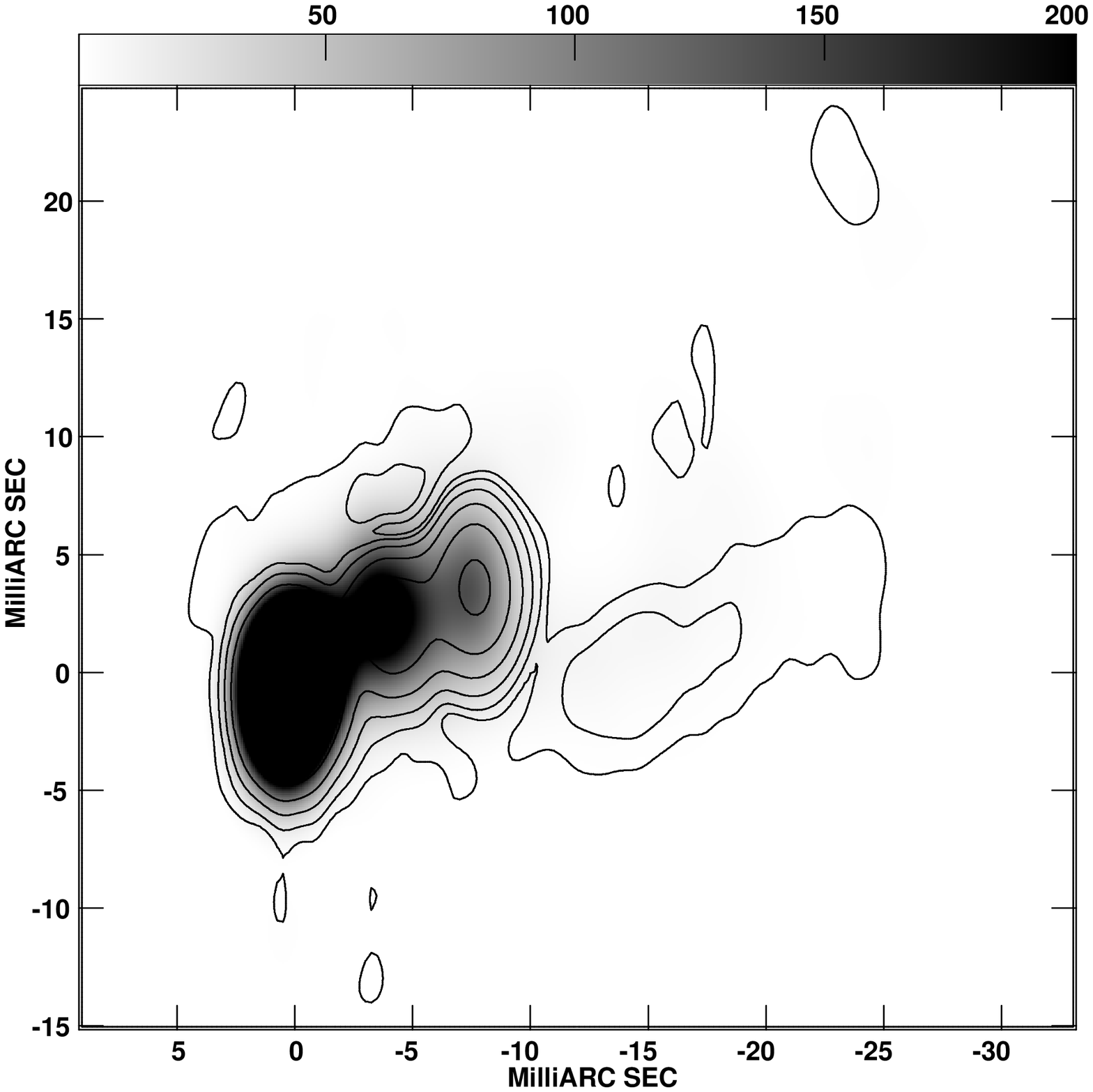}{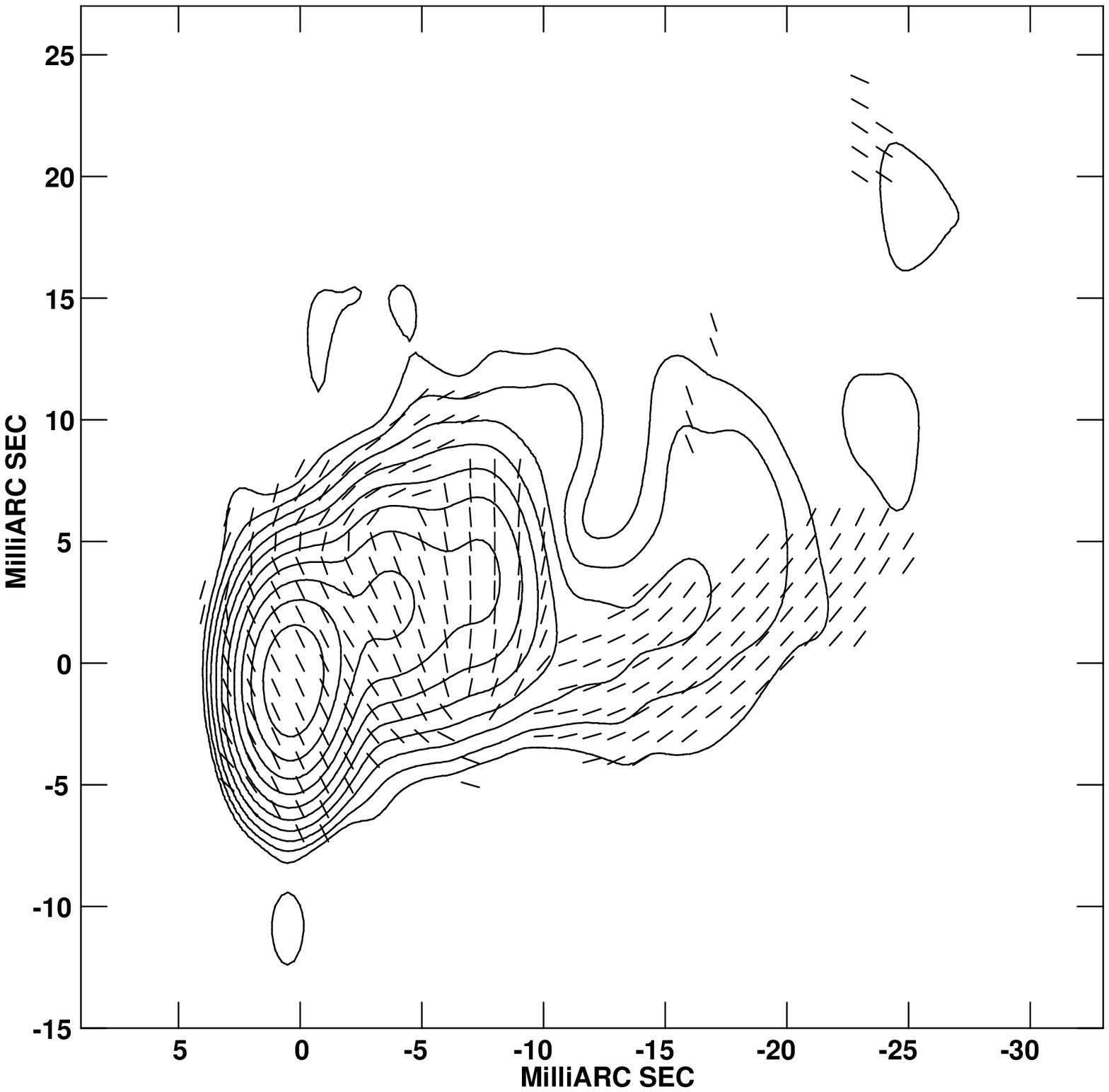}
\figcaption{Superposition of the total intensity and linear polarization
distributions of \src\/, epoch 1997.07, made with the VLBA$+$Y1 at 5~GHz. ($a$)
Every other linearly polarized intensity contour from Fig.~1$b$
over a gray-scale image of the total intensity (the scale at the top is $I$
in mJy~beam$^{-1}$); ($b$) every other total intensity contour from
Fig.~1$a$, plus tick marks showing the orientation $\theta =
\chi + \frac{\pi}{2}$ of the magnetic field in the source in those regions
where the linearly polarized intensity exceeds 0.5~mJy~beam$^{-1}$, the
lowest contour in ($a$).\label{fig:1055IP}}
\end{figure}

\clearpage

\begin{figure}
\plottwo{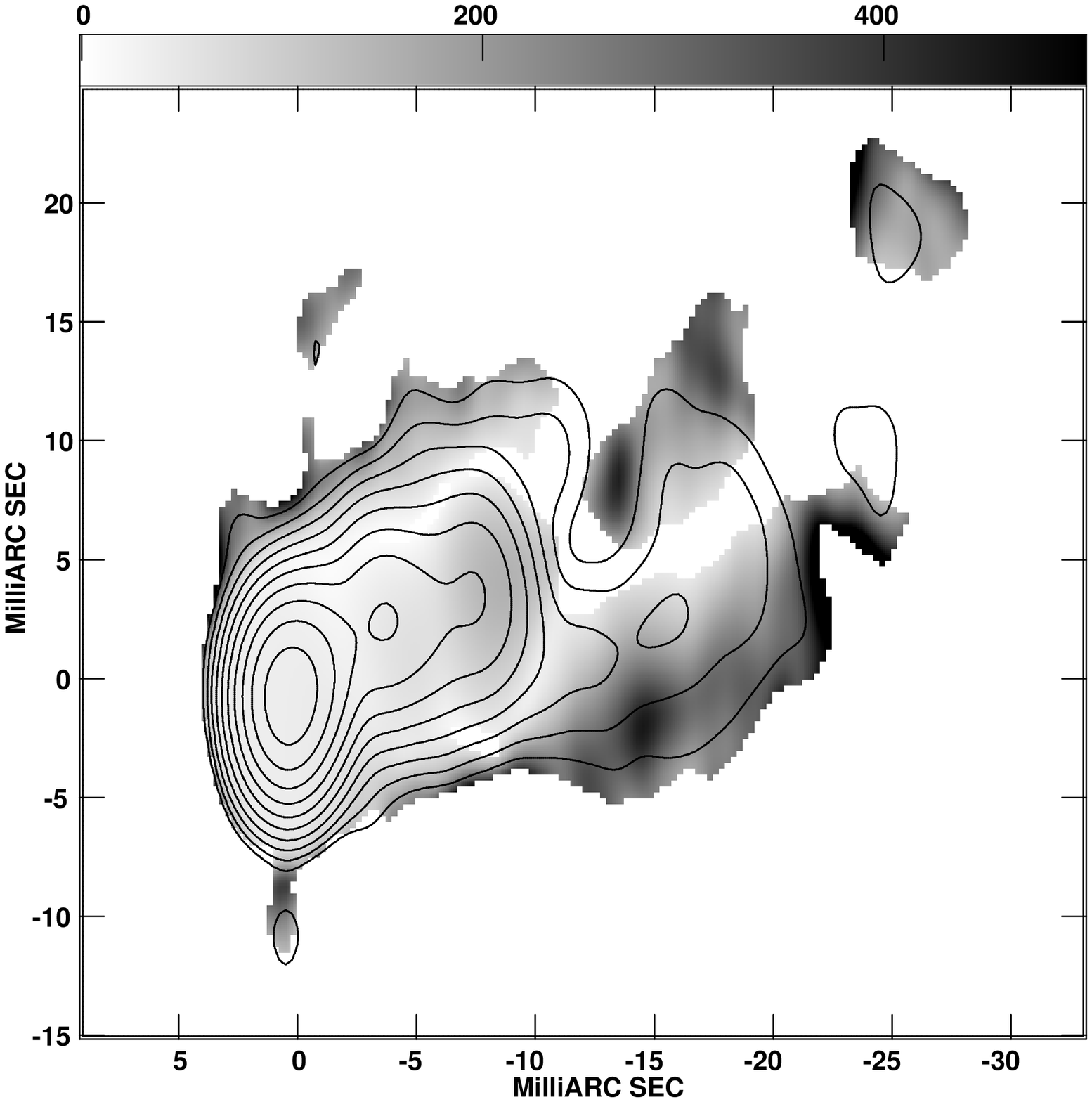}{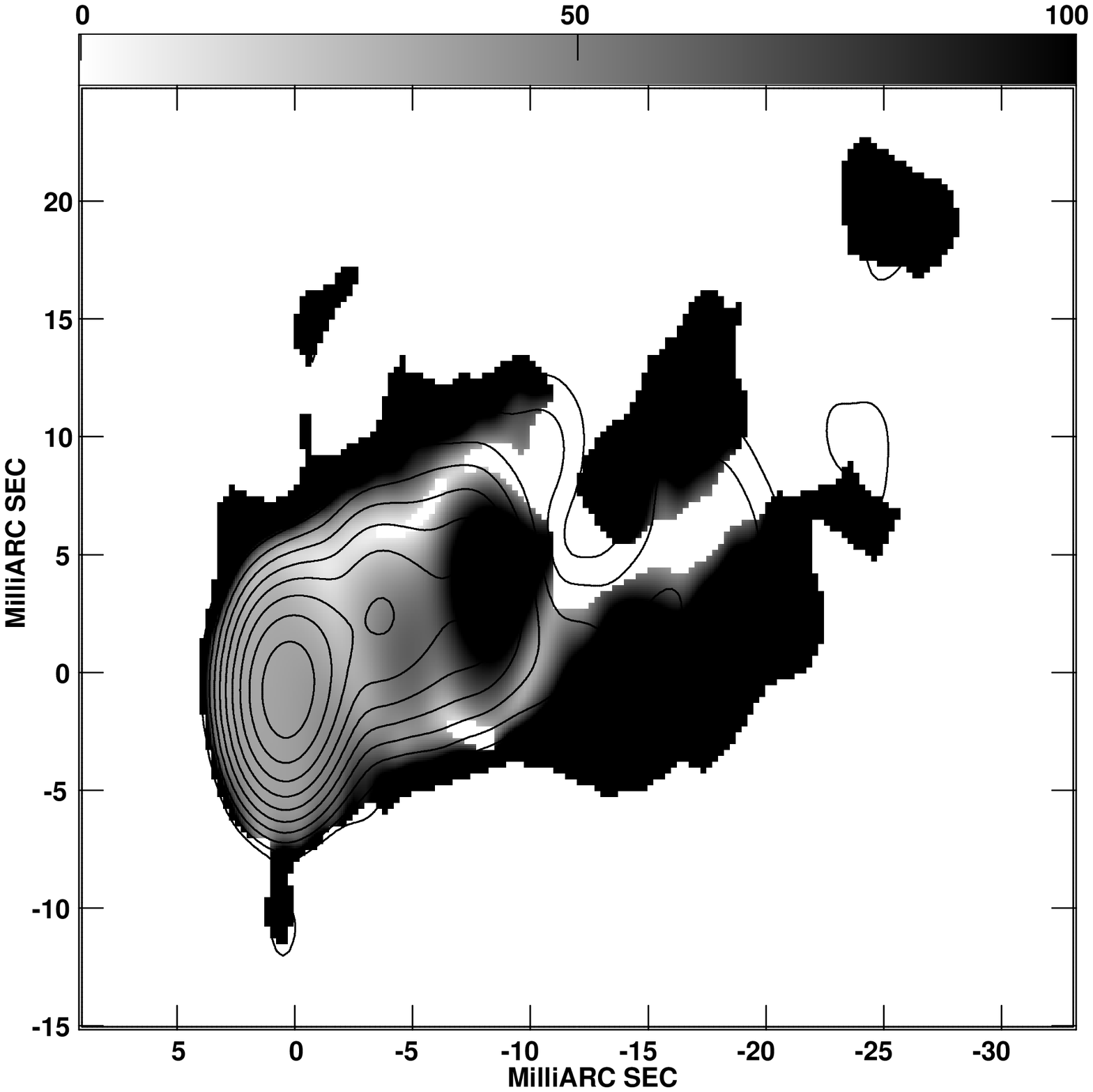}
\figcaption{Superposition of every other contour of total intensity $I$ from
Fig.~1$a$ over gray-scale for fractional linear polarization
$m$ in \src\ (the scales at top are $1000m$). ($a$) Polarization
saturated at $ m = 0.5$ (showing details of the boundary layer), and ($b$)
polarization saturated at $ m = 0.1$ (showing the spine).\label{fig:1055FPOL}}
\end{figure}

\clearpage

\begin{deluxetable}{cccccccccc}
\tablecolumns{10}
\footnotesize
\tablecaption{Jet Spine Components in \src}
\tablehead{
  \colhead{} & \colhead{} & \colhead{$r$} 
  & \colhead{$\theta$} & \colhead{$I$} 
  & \colhead{$m$} & \colhead{$\chi$} 
  & \colhead{Major~Axis} & \colhead{Minor~Axis} 
  & \colhead{$\phi$} \\
\colhead{Epoch} & \colhead{Component} & \colhead{(mas)}
  & \colhead{(deg)} & \colhead{(mJy)}
  & \colhead{(\%)} & \colhead{(deg)}
  & \colhead{(mas)} & \colhead{(mas)}
  & \colhead{(deg)} \\ 
\colhead{(1)} & \colhead{(2)} & \colhead{(3)}
  & \colhead{(4)} & \colhead{(5)}
  & \colhead{(6)} & \colhead{(7)}
  & \colhead{(8)} & \colhead{(9)}
  & \colhead{(10)} }

\startdata
1996.02 & D & \nodata & \nodata & 1190 & 2.8 & -37.5 & $<$ 0.35 & $<$ 0.15 & 0.0 \nl
 & C3 & 1.53 & -48.6 & 431 & 2.9 & -21.4 & 1.10 & $<$ 0.15 & -60.1 \nl
 & C2 & 5.26 & -53.0 & 510 & 5.6 & -71.8 & 3.38 & 2.76 & -82.8 \nl
 & C1 & 9.17 & -61.2 & 176 & 11.3 & 74.6 & 3.85 & 1.84 & 3.0 \nl
&&&&&&&&&\nl
1996.96 & D & \nodata & \nodata & 1350 & 3.1 & -73.8 & $<$ 0.35 & $<$ 0.15 & 36.8 \nl
 & C3 & 1.81 & -46.5 & 394 & 5.0 & -76.0 & 0.68 & $<$ 0.15 & -37.7 \nl
 & C2 & 5.30 & -53.5 & 420 & 4.7 & -72.8 & 2.70 & 2.52 & -42.4 \nl
 & C1 & 9.09 & -61.3 & 198 & 11.8 & 78.8 & 3.82 & 1.73 & 2.1 \nl
&&&&&&&&&\nl
1997.07 & D & \nodata & \nodata & 1590 & 3.5 & -73.4 & 0.46 & $< 0.15$ & -35.0 \nl
 & C3 & 1.79 & -46.8 & 435 & 3.9 & -72.4 & 0.79 & 0.19 & -34.4 \nl
 & C2 & 5.30 & -52.2 & 469 & 3.9 & -71.7 & 2.98 & 2.43 & 2.6 \nl
 & C1 & 9.08 & -61.3 & 276 & 10.4 & 78.1 & 4.51 & 2.17 & 2.6 \nl
\enddata
\tablecomments{Col.~(1): Epoch of observation; col.~(2): component name; col.~(3):
distance from
easternmost feature D; col.~(4): position angle of separation from D; col.~(5): total
intensity; col.~(6):
fractional linear polarization; col.~(7): orientation of the linear polarization position
angle; col.~(8): major axis of the model component; col.~(9): minor axis of the
model component;
and col.~(10): orientation of the major axis.}

\end{deluxetable}


\clearpage

\begin{thebibliography}{}

\bibitem[Aaron 1996]{A96} Aaron, S.~E. 1996, Ph.~D. Thesis, Brandeis Univ.

\bibitem[Antonucci 1993]{A93} Antonucci, R. 1993, Ann.~Rev.~Astron.~Astrophys.
31, 473

\bibitem[Baldwin, Wampler, \& Gaskel 1989]{BWG89} Baldwin, J.~A., Wampler, E.~J. \&
Gaskel, C.~M. 1989, ApJ, 338, 630

\bibitem[Bondi et al.\ 1996]{B96} Bondi, M., et al. 1996, A\&A, 308, 415

\bibitem[Cawthorne et al.\ 1993]{CWRG93} Cawthorne, T.~V., Wardle, J.~F.~C., Roberts,
D.~H., \& Gabuzda, D.~C. 1993, ApJ, 416, 519

\bibitem[Cotton 1993]{C93} Cotton, W.~D. 1993, AJ, 106, 1241

\bibitem[Falomo, Scarpa, \& Bersanelli 1994]{FSB94} Falomo, R., Scarpa, R., \&
Bersanelli, M. 1994, ApJS, 93, 125

\bibitem[Hughes, Aller, \& Aller 1985]{HAA85} Hughes, P.~A., Aller, H.~D., \& Aller,
M.~F. 1985, ApJ, 298, 301

\bibitem[Kim, Tribble, \& Kronberg 1991]{KTK91} Kim, K.~T., Tribble, P.~C. \&
Kronberg, P.~P. 1991, ApJ, 379, 80

\bibitem[Laing 1980]{L80} Laing, R.~A. 1980, MNRAS, 193, 439

\bibitem[Laing 1996]{L96} Laing, R.~A. 1996, In  Energy Transport in Radio Galaxies
and Quasars, ed.\  Hardee, P.~E., Bridle, A.~H. \& Zensus, J.~A. (PASP Conf.\
Series, Vol.\ 100, Astronomical Society of the Pacific, San Francisco), p.~241

\bibitem[Lepp\"{a}nen, Zensus, \& Diamond 1995]{LZD95} Lepp\"{a}nen, K.~J., Zensus,
J.~A., \& Diamond, P.~J. 1995, AJ, 110, 2479

\bibitem[Morris et al.\ 1991]{M91} Morris, S.~L., Stocke, J.~T., Gioia, I.~M., Schild,
R.~E., Wolter, A.,  Maccacaro, T., \& Della Ceca, R. 1991, ApJ, 380, 49

\bibitem[Murphy, Browne, \& Perley 1993]{MBP93} Murphy, D., Browne, I., \& Perley,
R. 1993, MNRAS, 264, 298

\bibitem[Roberts, Wardle, \& Brown 1994]{RWB94} Roberts, D.~H., Wardle, J.~F.~C., \&
Brown, L.~F. 1994, ApJ, 427, 718

\bibitem[Romney et al.\ 1984]{R84} Romney, J., et al. 1984, A\&A, 135, 289

\bibitem[Swain, Bridle, \& Baum 1996]{SBB96} Swain, M.~R., Bridle, A.~H., \& Baum,
S.~A. 1996, In Energy Transport in Radio  Galaxies and Quasars, ed.\ Hardee, P.~E.,
Bridle, A.~H. \& Zensus, J.~A. (PASP Conf.\ Series, Vol.\ 100, Astronomical Society of
the Pacific, San  Francisco), p.~299

\bibitem[Wardle et al.\ 1994] {WCRB94} Wardle, J.~F.~C., Cawthorne, T.~V., Roberts,
D.~H., \& Brown, L.~F. 1994, ApJ, 437, 122

\bibitem[Wills \& Lynds 1978]{WL78} Wills, D., \& Lynds, R. 1978, ApJS, 36, 317

\end{thebibliography}
\end{document}